\documentclass[12pt]{article}
\usepackage{amsfonts,amsmath,amssymb}
\usepackage{graphicx}

\newfont{\twelvemsb}{msbm10 scaled\magstep1}
\newfont{\eightmsb}{msbm8}
\newfam\msbfam
\textfont\msbfam=\twelvemsb \scriptfont\msbfam=\eightmsb
\catcode`\@=11
\def\Bbb{\ifmmode\let\next\Bbb@\else
\def\next{\errmessage{Use \string\Bbb\space only in math mode}}\fi\next}
\def\Bbb@#1{{\fam\msbfam{{#1}}}}

\newcommand{\be}{\begin{equation}}
\newcommand{\ee}{\end{equation}}
\newcommand{\ba}{\begin{eqnarray}}
\newcommand{\ea}{\end{eqnarray}}

\begin{document}

\sloppy
\renewcommand{\thefootnote}{\fnsymbol{footnote}}
\newpage
\setcounter{page}{1} \vspace{0.7cm}
\vspace*{1cm}
\begin{center}
{\bf The generalised scaling function: a note}\\
\vspace{1.8cm} {\large Davide Fioravanti $^a$, Paolo Grinza $^b$ and
Marco Rossi $^c$
\footnote{E-mail: fioravanti@bo.infn.it, pgrinza@usc.es, rossi@cs.infn.it}}\\
\vspace{.5cm} $^a$ {\em Sezione INFN di Bologna, Dipartimento di Fisica, Universit\`a di Bologna, \\
Via Irnerio 46, Bologna, Italy} \\
\vspace{.3cm} $^b${\em Departamento de Fisica de Particulas,
Universidad de Santiago de Compostela, 15782 Santiago de
Compostela, Spain} \\
\vspace{.3cm} $^c${\em Dipartimento di Fisica dell'Universit\`a
della Calabria and INFN, Gruppo collegato di Cosenza, I-87036
Arcavacata di Rende, Cosenza, Italy}
\end{center}
\renewcommand{\thefootnote}{\arabic{footnote}}
\setcounter{footnote}{0}
\begin{abstract}
{\noindent A method for determining the generalised scaling
function(s) arising in the high spin behaviour of long operator
anomalous dimensions in the planar $sl(2)$ sector of ${\cal N}=4$
SYM is proposed. The all-order perturbative expansion around the
strong coupling is detailed for the prototypical third and fourth
scaling functions, showing the emergence of the $O(6)$ Non-Linear
Sigma Model mass-gap from different SYM 'mass' functions.
Remarkably, only the fourth one gains contribution from the non-BES
reducible densities and also shows up, as first, NLSM interaction
and specific model dependence. Finally, the computation of the
$n$-th generalised function is sketched and might be easily
finalised for checks versus the computations in the sigma model or
the complete string theory.}
\end{abstract}
\vspace{6cm}

\newpage

\section{Introduction}

The planar $sl(2)$ sector of ${\cal N}=4$ SYM contains local
composite operators of the form
\begin{equation}
{\mbox {Tr}} ({\cal D}^s {\cal Z}^L)+.... \, , \label {sl2op}
\end{equation}
where ${\cal D}$ is the (symmetrised, traceless) covariant
derivative acting in all possible ways on the $L$ bosonic fields
${\cal Z}$. The spin of these operators is $s$ and $L$ is the
so-called 'twist'. Moreover, this sector would be described --
thanks to the AdS/CFT correspondence \cite {MWGKP} -- by string
states on the $\text{AdS}_5\times\text{S}^5$ spacetime with
$\text{AdS}_5$ and $\text{S}^5$ charges $s$ and $L$, respectively.
In addition, as far as the one loop is concerned, the Bethe Ansatz
problem is equivalent to that of twist operators in QCD
\cite{LIP,BDM}\footnote{A deep reason for that may be that one loop
QCD still shows up conformal invariance, albeit integrability does
not seem to hold in full generality (for instance, it apparently
imposes aligned partonic helicities).}.

Proper superpositions of operators (\ref {sl2op}) have definite
anomalous dimension $\Delta$ depending on $L$, $s$ and the 't Hooft
coupling $\lambda = 8 \pi^2 g^2$:
\begin{equation}
\Delta = L+s+\gamma (g,s,L) \, , \label{Delta}
\end{equation}
where $\gamma (g,s,L)$ is the anomalous part. A great boost in the
evaluation of $\gamma (g,s,L)$ in another sector has come from the
discovery of integrability for the purely bosonic $so(6)$ operators
at one loop \cite{MZ}. Later on, this fact has been extended to the
whole theory and at all loops in the sense that, for instance, any
operator of the form (\ref{sl2op}) is associated to one solution of
some Bethe Ansatz-like equations and thus any anomalous dimension
becomes a function, basically the energy, of one solution \cite{BS,
ES, BES}. Nevertheless, along with this host of new results an
important limitation emerged as a by-product of the {\it on-shell}
($S$-matrix) Bethe Ansatz: as soon as the interaction reaches a
range greater than the chain length, then it becomes modified by
unpredicted {\it wrapping effects}. More precisely, the anomalous
dimension is {\it in general} correct up to the $L-1$ loop in the
convergent, perturbative expansion, i.e. up to the order $g^{2L-2}$.
Which in particular implies, -- fortunately for us, -- that the {\it
asymptotic} Bethe Ansatz should give the right result whenever the
subsequent limit (\ref{jlimit}) is applied, as in the expansion
(\ref{sud}).

In fact, an important large twist and high spin scaling may be
considered on both sides of the correspondence\footnote{Actually, in
string theory (semi-classical) calculations the $\lambda\rightarrow
+\infty$ limit needs consideration before any other (cf. for
instance \cite{FTT} and references therein), thus implying a
different limit order with respect to our gauge theory approach (cf.
below for more details).}:
\begin{equation}
s \rightarrow \infty \, , \quad L \rightarrow \infty \, , \quad
j=\frac {L}{\ln s}={\mbox {fixed}} \, . \label{jlimit}
\end{equation}
The relevance of this logarithmic scaling for the anomalous
dimension of long operators has been pointed out in \cite{BGK}
within the (one-loop) SYM theory and then in \cite{FTT} and
\cite{AM} within the string theory (strong coupling $g\gg 1$). In
fact, by describing the anomalous dimension through a non-linear
integral equation \cite{FRS} (like in other integrable theories
\cite{FMQR}), it has been recently confirmed the Sudakov leading
behaviour for $s\rightarrow +\infty$ \cite{AM, FTT}
\begin{equation}
\gamma (g,s,L)=f(g,j)\ln s + \dots \, , \label{sud}
\end{equation}
thus indeed generalising to all loops a result by
\cite{BGK}.
Actually, this statement was argued in \cite{FRS} by computing
iteratively the solution of some integral equations and then,
thereof, {\it the generalised scaling function}, $f(g,j)$ at the
first orders in $j$ and $g^2$: more precisely, the first orders in
$g^2$ have been computed for the first {\it generalised scaling
functions} $f_n(g)$, forming
\begin{equation}
f(g,j)=\sum _{n=0}^{\infty} f_n(g)j^n \, .
\end{equation}
As a by-product, the reasonable conjecture has been put forward that
the two-variable function $f(g,j)$ should be analytic (in $g$ for
fixed $j$ and in $j$ for fixed $g$). In \cite{BFR2} similar results
have been derived for what concerns the contribution beyond the
leading scaling function $f(g)=f_0(g)$, but with a modification
which has allowed to neglect the non-linearity for finite $L$ and to
end-up with one linear integral equation. The latter does not differ
from the BES one (which cover the case $j=0$, cf. \cite{BES}), but
for the inhomogeneous term, which is the sum of an integral on the
one loop root density, a hole contribution and a known function. In
this respect, we have reckoned interesting the analysis of the
next-to-leading-order (nlo) term -- still coming, for finite $L$,
from an asymptotic Bethe Ansatz --, as the leading order $f(g)$ has
been conjectured to be independent of $L$ or {\it universal}
\cite{BGK}, \cite{ES} .

For instance, this nlo term enjoys many attractive features, like
its form which contains up to the $L$-linear term, i.e.
$f_1(g)L+c(g)$. In particular, the linearity in $L$ and the
behaviour of the next terms with increasing powers (of $L$) has
furnished us the {\it inspiration} for the past \cite{FGR} and
present calculations, respectively. In this respect, we deem useful
to briefly introduce in the next section a suitable modification of
the aforementioned method such that it applies to the regime
(\ref{jlimit}) (still for any $g$ and $j$). In fact, a suitable
modification of this LIE has been already exploited and explored in
\cite{FGR} to derive still a LIE in the scaling (\ref{jlimit}) (for
any $g$ and $j$). Along this path, we want to determine here the
generalised scaling function $f(g,j)$ and also its {\it
constituents} $f_n(g)$ for all the values of $j$ and $g$, thus
interpolating from weak to strong coupling: in fact, we will see in
the following why the expansion around $j=0$ is suitable, efficient
and manageable. Specifically, we will compute $f_n(g)$ for
$n=2,3,4$, show that $f_2(g)=0$ and derive the strong coupling
behaviour of $f_3(g)$ and $f_4(g)$. In this respect, a crucial point
is the appearance in the calculation regarding $f_4(g)$ (and onwards
for $n>4$) of the higher root densities, along with the BES density,
which are not expressible in terms of the BES one. This novel
feature (i.e. essentially new (recursive) linear integral equations
(concerning the higher densities) beside the BES one) plays an
important r\^ole in the comparison with the $O(6)$ Non-Linear Sigma
Model (NLSM). For, at the leading large $g$ order, we will find
perfect agreement with the infrared (i.e. strong coupling) expansion
of the $O(6)$ Non-Linear Sigma Model (NLSM) \cite{DF}, thus
enforcing the ultraviolet (i.e. weak coupling) checks and
predictions by Alday and Maldacena \cite{AM}. The IR results, on the
one hand, explore the other regime of the asymptotically free NLSM,
which develops a mass gap, on the other yield a convergent series
(for $f(g,j)$, in this limit, as a function of $j$ and $\forall g$
in the coefficients $f_n(g)$) \cite{DF}. This feature, the
convergence, likely extends to the full SYM theory, so bearing new
interest to this expansion. We would also remark here, but we will
discuss more deeply this point afterwards, that the agreement for
$f_4(g)$ (and onwards for $n>4$) is highly non-trivial since it
involves the specific interacting theory of the $O(6)$ NLSM.
Finally, the ideology of our method for computing the $n$-th
generalised scaling function $f_n(g)$ is exposed and shown to be
easily implementable by using analytical and numerical techniques.

Incidentally, in the subsequent section we will also briefly
understand the corrections (represented by dots) to the leading
Sudakov factor (\ref{sud}), as they are indeed formed by the terms
like $c(g)$ in the aforementioned nlo term.

{\bf Note added:} An interesting paper \cite{BB} appears today in
the web archives. It seems to have some goals and equations similar
to ours, although coming more directly from the approach \cite{FRS},
and giving the leading strong coupling behaviour of $f_3(g)$ in
agreement with us and confirming $f_2(g)=0$ as in \cite{FRS}, and
$f_1(g)$ as in \cite{FGR}. Nothing explicit we can find in it about
$f_4(g)$ and a systematic plan for all the other $f_n(g)$.

\section{Computing the generalised scaling functions}
\setcounter{equation}{0}

In the framework of integrability in ${\cal N}=4$ SYM, we have been
finding useful \cite{NOI} to re-write the Bethe equations as a
unique non-linear integral equation \cite {FMQR} \footnote{Only in
case of nested Bethe Ansatz we may have more than one equation, but
always in small number.}. In particular, we have developed (starting
in \cite{BFR}) a new technique in order to cope with the frequent
(in SYM theory) case where the Bethe roots are confined within a
finite interval \cite{BFR2}. The non-linear integral equation
regarding the $sl(2)$ sector involves two functions $F(u)$ and
$G(u,v)$, both satisfying linear equations. It is convenient to
split $F(u)$ into its one-loop and higher than one loop
contributions $F_0(u)$ and $F^H(u)$ respectively, and define the
quantities $\sigma _0(u)=\frac {d}{du} F_0(u)$ and $\sigma _H(u)
=\frac {d}{du} F^H(u)$, which, in the high spin limit, have the
meaning of one loop and higher than one loop density of Bethe roots
and holes, respectively.

It is important to stress that in this paper we will follow the
route initiated in \cite{BFR2}, and further developed in \cite{FGR}
where the (strong coupling) behaviour of the first constituent
$f_1(g)$ was studied in full detail. In this perspective, one of the
main aims of the present paper is to show how we can study the
generalised scaling function at \emph{all orders} in $j$. In other
words, we will show how to write down {\it recursive} linear
integral equations for infinite many densities, $\sigma_H^{(n)}$, \,
$n=0,1,2,\dots$ (the first one being the famous BES density
\cite{BES}), each one yielding $f_n(g)$, in a way which
systematically furnishes the $n$-th one, provided those
before\footnote{Starting, actually, from $n-3$ backwards.} are {\it
known}. Indeed, this is the main consequence of a suitable
manipulation of the initial integral equation for the full density
to end up with a recursive system of Fredholm (II type) linear
integral equations, all with the same kernel (the BES one), but
different inhomogeneous terms. The latter are the only parts entered
by the previous densities and are thus responsible for the recursive
derivation and solution. Eventually, we shall remark that this
structure has already appeared in \cite{FGR}, allowing us to study
$f_1(g)$: in fact, we did not even need the knowledge of the BES
density $\sigma_H^{(0)}$, thanks to the general rule that only
$\sigma_H^{(m)}$, with $m\leq n-3$, enter the inhomogeneous term
regarding $\sigma_H^{(n)}$.

Let us now analyse in detail the linear integral equations which are
satisfied by the one and higher loop densities and in terms of which
we may write the anomalous dimension.

For the time being, we want to extend to the scaling situation
(\ref{jlimit}) the procedure by \cite{BFR2, FGR} for computing the
one loop observables (i.e. the physical quantities depending on the
one loop density $\sigma_0(v)$). The integration on the finite range
of roots,  $(-b_0,b_0)$, may be computed via an infinite
integration\footnote{The parameter $b_0$ depends on the total number
of roots $s$ through the integral on $\sigma_0(v)$ (the
normalisation condition) which counts them (cf. \cite{BFR2}).}
\begin{equation}
\int _{-b_0}^{b_0} dv f(v) \sigma _0(v)= \int _{-\infty}^{\infty} dv
f(v) \sigma _0^s(v) +O(({\ln s})^{-\infty}) \, , \label{1loop}
\end{equation}
where the Fourier transform of the {\it effective density} $\sigma
_0^s(v)$ obeys the integral equation
\begin{equation}
\hat \sigma _0^s(k)=-4\pi \frac {\frac {L}{2}-e^{-\frac {|k|}{2}}
\cos \frac {ks}{\sqrt {2}}} {2\sinh \frac {|k|}{2}}-\frac {e^{-\frac
{|k|}{2}}}{2\sinh \frac {|k|}{2}} \int  _{-\infty}^{\infty} du
e^{iku} \chi _{c_0} (u) \sigma _0^s(u)-4\pi \delta (k) \ln 2 \, ,
\label {sigma0}
\end{equation}
with $(-c_0, c_0)$ the interval spanned by the internal holes
\cite{BFR2}. This simply means
\begin{equation}
\int  _{-\infty }^{\infty} du \chi _{c_0} (u)\sigma
_0^s(u)=-2\pi(L-2) + O(({\ln s})^{-\infty}) \, , \label {sigma0cond}
\end{equation}
with the interval function $\chi_{c_0}(u)$ equal to $1$ if $-c_0\leq
u \leq c_0$ (here the internal holes concentrate), and $0$ otherwise
(no internal holes outside). Within the double limit (\ref{jlimit}),
the above remainders are $O(({\ln s})^{-\infty})$ \footnote{$O(({\ln
s})^{-\infty})$ means a remainder which goes to zero faster that any
inverse power of $\ln s$: $\lim\limits_{s \rightarrow \infty} (\ln
s)^k O((\ln s)^{-\infty}) =0, \forall \, k>0$.} and are exactly
given by the non-linear integrals in \cite{BFR2} we have here
dropped out.

Now, we ought to briefly comment on the corrections of the leading
Sudakov formula (\ref{sud}), namely the dots in there. We have
deduced in \cite{BFR2} a linear integral equation suitable to detail
the appearance of a constant term $O((\ln s)^0)$, denoted in the
introduction as $f_1(g)L+c(g)$\footnote{At small $g$, the same term
has been derived in \cite{FRS} by evaluating at leading order the
non-linear integrals of a {\it traditional} NLIE approach
\cite{FMQR}.}. Concerning this, the first addendum is the
generalised scaling function at first $j$-order \cite{FGR}, whilst
the second one gives rise to a $O((\ln s)^0)$ correction to the
leading Sudakov scaling, i.e. the $j^0$ order of some $f^{(0)}(g,j)$
in the expansion
\begin{equation}
\gamma (g,s,L)=f(g,j)\ln s + f^{(0)}(g,j) +  \dots \, \, .
\label{corrsud}
\end{equation}
Now, similarly we may imagine that the dots should be inverse
integer powers of $\ln s$, with coefficients, at each power,
depending on $g$ and $j$\footnote{Naturally the subsequent expansion
(\ref{c0series}) and its higher loop analogue ought to be amended
accordingly.}. Very interestingly, these power-like corrections to
the leading (\ref{sud}) seem to come out truly from our systematic
expansion of the one-loop density, -- satisfying a linear integral
equation --, and its subsequent insertion into the inhomogeneous
term of the higher loop linear integral equation (i.e.
(\ref{sigmaeq})). In fact, the last $\delta$-term in (\ref{sigma0})
contributes, for instance, to $f^{(0)}(g,j)$, namely to the
mentioned $c(g)$, whilst it does not to $f(g,j)$\footnote{The other
part of $c(g)$ derives from the rest of the equation, i.e. the next
equation (\ref{hatsigma0}), but it will be neglected in the
following.}. In other words, all these terms seem to be controlled
by linear equations\footnote{This is similarly true for the linear
integral formula of the anomalous dimension.}, after neglecting the
non-linear integrals. Eventually, there might also be a possible
non-analytic (in the variable $\ln s$) correction, $O(({\ln
s})^{-\infty})$. Yet, the $O(({\ln s})^{-\infty})$ term would face
two problems: on the one side it is determined by cumbersome
non-linear integrals which deny the density (linear) treatment, on
the other side it may well become affected by the non-asymptotic
phenomenon of wrapping.

Therefore, as we will be constraining ourselves to the leading
Sudakov factor $f(g,j)$, we can neglect the aforementioned
$\delta$-term and we may, for convenience' sake, Fourier-transform
the equations (\ref{sigma0}) and (\ref{sigma0cond}), respectively:
\begin{eqnarray}
\hat \sigma _0^s(k)=-4\pi \frac {\frac {L}{2}-e^{-\frac {|k|}{2}}
\cos \frac {ks}{\sqrt {2}}}
{2\sinh \frac {|k|}{2}}
&-& \frac {e^{-\frac {|k|}{2}}}{\sinh \frac {|k|}{2}} \int
_{-\infty}^{\infty} \frac {dh}{2\pi} \hat \sigma _0^s(h) \frac {\sin
(k-h)c_0}{k-h}  \, , \label {hatsigma0} \\
2 \int  _{-\infty }^{\infty} \frac {dk}{2\pi} \hat \sigma _0^s(k)
\frac {\sin kc_0}{k} &=& -2\pi(L-2) \, .
\label {hatsigma0cond}
\end{eqnarray}

These final relations characterise the one-loop theory in the high
spin and large twist limit (\ref{jlimit}) and
need to be solved together in order to give $\hat
\sigma _0^s(k)$ and $c_0$. In specific, we may solve the first one,
a Fredholm (II type) integral equation,
by Neumann-Liouville (recursive) series and then expand usefully $c_0(j)$
according to
\begin{equation}
c_0(j)=\sum _{n=1}^{\infty} c_0^{(n)}j^n \, . \label {c0series}
\end{equation}
Already the first step of this procedure (the approximation by the
inhomogeneous term) yields the first two coefficients of this series
\begin{equation}
c_0^{(1)}=\frac {\pi}{4} \, , \quad c_0^{(2)}=-\frac {\pi}{4}\ln 2
\, .
\end{equation}

Regarding the high spin behaviour of long operators, the approach to
the higher than one loop density, $\sigma _{H}(u)$, moves along
similar lines. Actually, the knowledge of $\sigma _0^s(u)$ concurs
to find it as the solution of the linear integral equation
(discarding the following $O(({\ln s})^{-\infty})$)
\begin{eqnarray}
\sigma _{H}(u)&=& -iL \frac {d}{du} \ln \left ( \frac {1+\frac {g^2}
{2{x^-(u)}^2}}{1+\frac {g^2}{2{x^+(u)}^2}} \right ) + \frac {i}{\pi}
\int _{-\infty}^{\infty} dv  \chi _{c} (v)\Bigl [\frac {d}{du} \ln
\left ( \frac {1-\frac {g^2}{2x^+(u)x^-(v)} }{1-\frac
{g^2}{2x^-(u)x^+(v)}} \right )
+ \nonumber \\
&+& i \frac {d}{du}\theta (u,v)  +i \frac {1}{1+(u-v)^2}\Bigr
][\sigma _0^s(v) +\sigma _H(v)]
\nonumber \\
&-&  \frac {i}{\pi} \int _{-\infty}^{\infty} dv \frac {d}{du}\left [
\ln \left ( \frac {1-\frac {g^2}{2x^+(u)x^-(v)} }{1-\frac
{g^2}{2x^-(u)x^+(v)}} \right )
+i \theta (u,v) \right ] \sigma _0^s(v)+  \label{sigmaeq} \\
&+&\int _{-\infty}^{+\infty} \frac {dv}{\pi} \frac {1}{1+(u-v)^2}\sigma _{H}(v) +\int _{-\infty}^{\infty} \frac {dv}{\pi} \chi _{c_0} (v) \frac {1}{1+(u-v)^2}\sigma _{0}^s(v)- \nonumber \\
&-& \frac {i}{\pi} \int _{-\infty}^{+\infty} dv \frac {d}{du} \left
[ \ln \left ( \frac {1-\frac {g^2}{2x^+(u)x^-(v)} }{1- \frac
{g^2}{2x^-(u)x^+(v)}} \right )+i \theta (u,v) \right ]  \sigma
_{H}(v) +O(({\ln s})^{-\infty})\, , \nonumber
\end{eqnarray}
constrained by the two conditions on $c$ and $c_0$ respectively
\begin{eqnarray}
&& \int  _{-\infty}^{\infty} du \chi  _c (u) [\sigma _0^s(u)+\sigma
_{H}(u)] =-2\pi(L-2) \, , \nonumber \\
\label {normale2} \\
&& \int  _{-\infty }^{\infty} du \chi _{c_0} (u)\sigma
_0^s(u)=-2\pi(L-2)  \, , \nonumber
\end{eqnarray}
with the former fixing the interval $(-c,c)$ where the all-loop
internal holes concentrate as well as the latter, equivalent to
(\ref{hatsigma0cond}), determines the range of the one-loop internal holes
\footnote {The function $\theta (u,v)$ appearing in (\ref
{sigmaeq}) is the 'dressing factor', introduced on the string side
in the second of \cite {BS} and finalised as meromorphic function of
the coupling constant in \cite {BES}.}.

As in the one loop case, in order to gain a better insight it is
convenient to re-write (\ref{sigmaeq}) in terms of Fourier
transforms (upon neglecting $O((\ln s)^{-\infty})$)
\begin{eqnarray}
&&\hat \sigma _H(k)=\pi L \frac {1-J_0({\sqrt {2}}gk)}{\sinh \frac
{|k|}{2}}
+   \nonumber \\
&+& \frac {1}{\sinh \frac {|k|}{2}} \int _{-\infty }^{\infty} \frac
{dh}{|h|} \Bigl [ \sum _{r=1}^{\infty} r (-1)^{r+1}J_r({\sqrt
{2}}gk) J_r({\sqrt {2}}gh)\frac {1-{\mbox {sgn}}(kh)}{2}
e^{-\frac {|h|}{2}} + \nonumber \\
&+&{\mbox {sgn}} (h) \sum _{r=2}^{\infty}\sum _{\nu =0}^{\infty}
c_{r,r+1+2\nu}(-1)^{r+\nu}e^{-\frac {|h|}{2}} \Bigl (
J_{r-1}({\sqrt {2}}gk) J_{r+2\nu}({\sqrt {2}}gh)- \label {sigmaeq2} \\
&-& J_{r-1}({\sqrt {2}}gh) J_{r+2\nu}({\sqrt {2}}gk)\Bigr ) \Bigr ]
\Bigl [\hat \sigma _0^s(h)+ \hat \sigma _H(h)-\int  _{-\infty
}^{\infty} \frac {dp}{2\pi} \left (\hat \sigma _0^s(p)+\hat \sigma
_H(p) \right ) 2 \frac {\sin (h-p)c} {h-p} \Bigr]
- \nonumber \\
&-&   \frac {e^{-\frac {|k|}{2}}}{\sinh \frac {|k|}{2}} \int
_{-\infty }^{\infty} \frac {dp}{2\pi} \left (\hat \sigma
_0^s(p)+\hat \sigma _H(p) \right ) \frac {\sin (k-p)c} {k-p} + \frac
{e^{-\frac {|k|}{2}}}{\sinh \frac {|k|}{2}} \int  _{-\infty
}^{\infty} \frac {dp}{2\pi} \hat \sigma _0^s(p) \frac {\sin
(k-p)c_0} {k-p}  \, , \nonumber
\end{eqnarray}
as well as the previous conditions on the internal holes
distributions
\begin{eqnarray}
2 \int  _{-\infty }^{\infty} \frac {dk}{2\pi} \hat \sigma _0^s(k)
\frac {\sin kc_0}{k} &=&-2\pi(L-2)  \, , \nonumber \\
\label{normale} \\
2 \int  _{-\infty }^{\infty} \frac {dk}{2\pi} [\hat \sigma
_0^s(k)+\hat \sigma _H(k)] \frac {\sin kc}{k} &=&-2\pi(L-2)  \, .
\nonumber
\end{eqnarray}

Eventually, the three equations (\ref{sigmaeq2}) and (\ref{normale})
form the starting point of our all loop analysis, generalising the
one loop statement. In fact, we may write the anomalous dimension
(in the usual limit (\ref{jlimit})) as
\begin{eqnarray}
\gamma (g,s,L)&=&-g^2\int _{-b_0}^{b_0}\frac {dv}{2\pi} \left [\frac
{i}{x^+(v)}-\frac {i}{x^-(v)}\right ] \sigma_0(v)+
\nonumber \\
&+&g^2 \int _{-\infty}^{+\infty} \frac {dv}{2\pi} \chi _c(v) \left [
\frac {i}{x^+(v)}-\frac {i}{x^-(v)}\right ] [  \sigma_0(v)+
\sigma _H(v) ] -  \label{egs} \\
&-& g^2\int _{-\infty}^{+\infty}\frac {dv}{2\pi} \left [\frac
{i}{x^+(v)}-\frac {i}{x^-(v)}\right ] \sigma_H(v) + O(({\ln
s})^{-\infty}) \, , \nonumber
\end{eqnarray}
and realise, upon comparing this with (\ref {sigmaeq2}), that
\begin{equation}
\gamma (g,s,L)=\frac {1}{\pi} \lim _{k\rightarrow 0}\hat \sigma
_H(k) + O(({\ln s})^{-\infty}) \label{egs2}\, .
\end{equation}
This relation extends the validity of its Kotikov-Lipatov analogue
\cite{KL} (valid at the leading order $\ln s$) to all finite orders
$({\ln s})^{-n}$, $n\in {\mathbb N}$. A statement of this type has
been already noted by \cite{FRS} in connection with $f(g,j)$ only
\footnote{For instance, we verified this for $f^{(0)}(g,j)$, albeit
we do not show it here.} and their equations, which have a different
form with respect to our (\ref{sigmaeq2}), though.

Now, a very crucial point of our construction enters the stage. The
key equation (\ref{sigmaeq2}) concerning the all loop density
$\sigma _H(u)$ and also the conditions (\ref{normale}) show up a
deep, even conceptual, difficulty to be solved for generic values of
the parameters $g$ and $j$ (or $s$, $L$). Yet, finding the solution
becomes easier if we think of it and of the internal holes boundary
$c(j)$ as an expansion in (non-negative) powers of $j$
\begin{equation}
\sigma _H(u)=[\sum _{n=0}^{\infty} \sigma _H^{(n)}(u)j^n]\ln s  \, , \quad c(j)=\sum _{n=1}^{\infty}
c^{(n)}j^n \, . \label{csigmaexp}
\end{equation}
For in this way the equation (\ref{sigmaeq2}) breaks down into a
{\it recursive} linear system of integral equations for the {\it
densities} $\sigma_H^{(n)}$, each one yielding $f_n(g)$ according to
(\ref{egs2}):
\begin{equation}
f_n(g)=\frac{1}{\pi}\lim _{k\rightarrow 0} \hat \sigma _H^{(n)}(k)
\label{f_n}.
\end{equation}
Importantly, each $\sigma_H^{(n)}$ is governed by a Fredholm (II
type) linear integral equation with always the same kernel (indeed
the BES one \cite{BES}), but a different inhomogeneous term. This
contains a part equal for any $n$ and given by the one loop density
(and an additional known function) and another part which instead
varies, but is still expressible - after using the conditions (\ref
{normale}) - in terms of only $\sigma_H^{(m)}$, with $m\leq n-3$.
Therefore, a recursive solution procedure can be contrived, at least
in principle. In \cite{FGR}, the first step of the recursive
structure has already appeared, without any reference to the BES
density $\sigma_H^{(0)}$, and allowed us to analyse $f_1(g)$. From a
physical point of view, the expansion in $j$ means that we are
focusing our attention on the regime $j\ll g$, with fixed coupling
$g$. Unfortunately, semi-classical string results concern a
different regime where $g$ is very large and {\it then} $j$ scales,
instead, accordingly, i.e. $j\sim g$ (in fact, the scaling variable
$y=j/g$ is fixed) \cite{FTT}. Nevertheless, our regime is suitable
for comparison with the string reduction to the $O(6)$ NLSM which
happens for $j\ll g$ as brilliantly conjectured by Alday and
Maldacena on a geometric ground \cite{AM}. These authors have also
checked their statement by using the perturbative renormalisation
procedure in the UV regime, where the mass-gap $m(g)$ is much
smaller than the energy density $j$, i.e. $j\gg m(g)$. On the
contrary, our kind of expansion obliges us to consider $j$ small and
then to consider the other NLSM regime, the IR one $j\ll m(g)$. In
this case, the perturbative field theory methods are much less
effective, but fortunately the Bethe Ansatz formulation, based on
the $S$-matrix, furnishes easily the necessary data still in a
systematic way \cite{DF}. The first check of the mass gap arising
has been confirmed for the first time in \cite{FGR} by computing
$f_1(g)$, while it could not be shown exactly in the strong coupling
behaviour of the cusp anomalous dimension $f_0(g)=f(g)$ because of
the UV cut-off effects on the NLSM vacuum energy. Instead, $f_1(g)$
gives rise to the simplest manifestation of the mass-gap and the
other scaling functions $f_n(g)$ -- as we are going to see --
present other different functions (of $g$), which all coincide with
the mass-gap in the strong coupling regime.

In specific, we will illustrate explicitly our general construction
in the following section giving as prototypical example the first
fully general case, i.e. that concerning $f_4(g)$. As a matter of
fact, this is the lowest order at which this computation is not
reducible to the {\it knowledge} of the BES density $\sigma_H^{(0)}$
(at least by us and by now), as it shows up the explicit appearance
of the higher order density $\sigma_H^{(1)}$, in the inhomogeneous
term. Curiously, it is also the lowest order at which the comparison
with the $O(6)$ NLSM becomes constraining, since right at this order
the specific interaction starts appearing. In other words, up to the
order $j^3$, the strong coupling expansion obtained for the $O(6)$
NLSM only relies on the non-interacting Fermi gas theory, which has
no information about the interaction and is in some sense
"universal", being , in particular, the same for any $N$ of the
$O(N)$. In other words, the exact leading order of $f_n(g)$ -- as it
comes out from the gauge theory -- is in perfect agreement with the
corresponding one as worked out in \cite{DF} within the $O(6)$ NLSM.
In the following, we have decided to illustrate the computation up
to the first general example, $f_4(g)$, which teaches us the general
procedure. Nevertheless, all the details and important subtleties
about the latter need to be given extensively in a separate
publication; for this reason we have simplified the illustration and
have limited ourselves to an effective numerical presentation of the
leading non-perturbative terms, which are responsible for the
mass-gap. Yet, the different physical origin of the $f_n(g)$ will
clearly prove that the complete NLSM allows (with its unique
mass-gap) only for their leading terms, though giving rise to an
impressive exact agreement.

\section{On the calculation of the generalised scaling functions: a sketch}
\setcounter{equation}{0}

In the present section we will sketch how to compute the
coefficients of the expansion (\ref{csigmaexp}), so that, by means
of the equality (\ref{egs2}), we can gain all the generalised
scaling functions $f_n(g)$. In this respect, computing the first
orders in the series for $c(j)$ (\ref{csigmaexp}) is instructive to
see how the procedure works and  will turn out useful for expanding
the subsequent forcing terms (cf. sub-section 3.3). The computation goes
along the similar lines of the one loop theory and yields
\begin{equation}
c^{(1)}=\frac {\pi}{4-\sigma _H^{(0)}(0)} \, , \quad c^{(2)}=-\pi
\frac {4\ln 2 -\sigma _H^{(1)}(0)}{[4-\sigma _H^{(0)}(0)]^2} \, .
\label{c1,c2}
\end{equation}

\subsection{The second generalised scaling function}

We now show that $\sigma _{H}^{(2)}(u)=0$, so that we obviously have
$f_2(g)=0$. Let us consider the r.h.s. of (\ref {sigmaeq}). The
first term is clearly proportional to $j\ln s$, so it does not
appear in the equation for $\sigma _{H}^{(2)}(u)$. The second and
the fifth term both have the form, with two different functions
$f(v)$,
\begin{equation}
\int  _{-\infty}^{+\infty} dv f(v)\sigma (v) \chi _{d} (v)= \int
_{-\infty}^{+\infty} \frac {dk}{2\pi} \hat f(k) \int
_{-\infty}^{+\infty} \frac {dp}{2\pi} \hat \sigma (p)
 2 \frac {\sin (k-p)d}{k-p} \, , \label {forc1}
\end{equation}
where $\sigma , d$ stand for $\sigma _0^s+\sigma _H, c$ if we
consider the second term and for $\sigma _0^s, c_0$ if we consider
the fifth term. Using the normalization condition
\begin{equation}
2 \int  _{-\infty }^{\infty} \frac {dp}{2\pi} \hat \sigma (p) \frac
{\sin pd}{p} =-2\pi L + O(({\ln s})^{-\infty}) \, ,
\end{equation}
one can show that
\begin{equation}
2 \int  _{-\infty}^{+\infty} \frac {dp}{2\pi} \hat \sigma (p)
 \frac {\sin (k-p)d}{k-p}=[-2\pi j + O(d^3)]\ln s  \, .
\end{equation}
Since $d$ starts from order $j$ in its expansion, the second and the
fifth term in the r.h.s. of (\ref {sigmaeq}) lack of the order
$j^2\ln s $ terms in their expansion. The same reasoning, applied to
the second term in the rhs of (\ref {hatsigma0}) - the one
containing the integral - implies that also this term lacks of the
order $j^2\ln s$. Therefore, the third term in the r.h.s. of (\ref
{sigmaeq}) is missing the quadratic order as well. It follows that
the equation for $\sigma _{H}^{(2)}(u)$ is
\begin{eqnarray}
\sigma _{H}^{(2)}(u)&=& \int _{-\infty}^{+\infty} \frac {dv}{\pi}
\frac {1}{1+(u-v)^2}
\sigma _{H}^{(2)}(v) \\
&-& \frac {i}{\pi} \int _{-\infty}^{+\infty} dv \frac {d}{du} \left
[ \ln \left ( \frac {1-\frac {g^2}{2x^+(u)x^-(v)} }{1- \frac
{g^2}{2x^-(u)x^+(v)}} \right )+i \theta (u,v) \right ] \sigma
_{H}^{(2)}(v) \, , \nonumber
\end{eqnarray}
whose solution is, of course, $\sigma _{H}^{(2)}(u)=0$. Therefore
$f_2(g)=0$.

\subsection{The third generalised scaling function}

In this sub-section, we concentrate our attention on the equation
for $\sigma _{H}^{(3)}(u)$. We find convenient to pass to the
Fourier transforms of all the involved quantities. Let us define the
even function
\begin{equation}
s^{(3)}(k)= \frac {2\sinh \frac {|k|}{2}}{2\pi  |k|}\hat \sigma
_H^{(3)}(k) \label {S3def} \, ,
\end{equation}
which has the property
\begin{equation}
\lim _{k\rightarrow 0}s^{(3)}(k)=\frac {1}{2}f_3(g) \, . \label
{sf_3}
\end{equation}
For future convenience, we will study the new function
\begin{equation}
S^{(3)}(k)=s^{(3)}(k)-\frac {\pi ^2}{6}|k| e^{-\frac {|k|}{2}}\left
[\frac {1}{16}-\frac {1}{[\sigma ^{(0)}(0)]^2}\right ] \, ,
\end{equation}
which depends on the all loops density of roots at (the zero-th)
order $\ln s$ (satisfying the BES equation \cite{BES}) at $u=0$
(real space)
\begin{equation}
\sigma ^{(0)}(0) = -4+\sigma ^{(0)}_H (0) \, ,
\end{equation}
(being the zero-th one-loop density in zero $\sigma ^{(0)}_0
(0)=-4$) and still keeps the key property (\ref{f_n}) in the form
\begin{equation}
\lim _{k\rightarrow 0}S^{(3)}(k)=\frac {1}{2}f_3(g) \, . \label
{Sf_3}
\end{equation}

Let us also introduce the functions
\begin{equation}
A^{(3)}_r(g)=\frac {\pi ^2 r}{6 [-4+\sigma ^{(0)}_H(0) ]^2}\int
_{0}^{+\infty}dh \, h \,  \frac {J_{r}({\sqrt {2}}gh)}{\sinh \frac
{h}{2}} \, , \label{A3}
\end{equation}
and restrict our analysis within the domain $k\geq 0$. Upon
expanding in series of Bessel functions (the so-called Neumann's
expansion) as
\begin{equation}
S^{(3)}(k)=\sum _{p=1}^{\infty}S^{(3)}_{2p}(g)\frac {J_{2p}({\sqrt
{2}}gk)}{k}+\sum _{p=1}^{\infty}S^{(3)}_{2p-1}(g)\frac
{J_{2p-1}({\sqrt {2}}gk)}{k} \, , \label{Neu}
\end{equation}
introducing the quantities\footnote{Their appearance comes out, also
but not only, from the coefficients of the dressing phase
$c_{r,s}(g)=2 \cos \left [ \frac {\pi}{2}(s-r-1)\right ] (r-1)(s-1)
Z_{r-1,s-1}(g)$ as in \cite{BES}.}
\begin{equation}
Z_{n,m}(g)=\int _{0}^{\infty} \frac {dh}{h} \frac {J_n({\sqrt
{2}}gh)J_m({\sqrt {2}}gh)}{e^h-1} \, ,
\end{equation}
we finally obtain the following infinite system of linear equations:
\begin{eqnarray}
S^{(3)}_{2p}(g)&=&A^{(3)}_{2p}(g)-4p\sum
_{m=1}^{\infty}Z_{2p,2m}(g)S^{(3)}_{2m}(g)+
4p\sum _{m=1}^{\infty}Z_{2p,2m-1}(g)S^{(3)}_{2m-1}(g)  \nonumber \\
S^{(3)}_{2p-1}(g)&=&A^{(3)}_{2p-1}(g)-2(2p-1)\sum
_{m=1}^{\infty}Z_{2p-1,2m}(g)
S^{(3)}_{2m}(g)-\label {Seq2} \\
&-& 2(2p-1)\sum _{m=1}^{\infty}Z_{2p-1,2m-1}(g)S^{(3)}_{2m-1}(g)
\nonumber \, .
\end{eqnarray}
The Neumann's decomposition (\ref{Neu}) reveals itself useful also
thanks to the special r\^ole of the first component in the key
relation
\begin{equation}
f_3(g)={\sqrt {2}}g S^{(3)}_{1}(g), \label{f_3,2}
\end{equation}
easily derived from (\ref{Sf_3}).

As a first approach, if we forget the non-perturbative contributions
around $g=+\infty$\footnote{Sometimes these are improperly called
{\it non-analytic}, in the sense that they have this character,
although the following series is (as a consequence!) not convergent
at all.}, we may seek for the solution to the equations (\ref{Seq2})
as an asymptotic series
\begin{equation}
S_{2m}^{(3)}(g)=\sum _{n=1}^{\infty } \frac
{S_{2m}^{(3,2n)}}{g^{2n}} \, , \quad S_{2m-1}^{(3)}(g)=\sum
_{n=2}^{\infty } \frac {S_{2m-1}^{(3,2n-1)}}{g^{2n-1}} \, .\label
{sexpan}
\end{equation}
The first two coefficients of (\ref{sexpan}) at the orders $1/g$ and
$1/g^2$ have the same structure as the corresponding ones in the
case of the first constituent $f_1(g)$ \cite{FGR}. Therefore, we are
naturally led to make the same Ansatz on the form of the
coefficients entering (\ref {sexpan}):
\begin{eqnarray}
&& S_{2m}^{(3,2n)}=2m \frac {\Gamma (m+n)}{\Gamma
(m-n+1)}(-1)^{1+n}b_{2n}^{(3)} \, , \quad
  n \geq 1 \, , \quad m \geq 1 \, , \nonumber \\
&& \label {guess} \\
&& S_{2m-1}^{(3,2n-1)}=(2m-1)\frac {\Gamma (m+n-1)}{\Gamma
(m-n+1)}(-1)^n b_{2n-1}^{(3)} \, , \quad n \geq 2 \, , \quad m \geq
2 \, . \nonumber
\end{eqnarray}
Analogously to the case of the first constituent, this Ansatz
implies that $S_{2m-1}^{(3,2n-1)}$ and $S_{2m}^{(3,2n)}$ are
different from zero only if $n\leq m$. The coefficients $b_n^{(3)}$
are unknown and they are determined by inserting the Ansatz
(\ref{sexpan}, \ref{guess}) into the equations (\ref {Seq2}).

In this way, implementing the elegant asymptotic expansion
\begin{equation}
A^{(3)}_r(g)=\frac {\pi ^2 r}{6 [\sigma ^{(0)}(0) ]^2} 2 \sum _{k=0}
^{\infty} \frac {2^{\frac {1}{2}-k}-2^{k-\frac {1}{2}}}{(2k)!}B_{2k}
\frac {\Gamma \left (\frac {1}{2}+\frac {r}{2}+k\right)}
 {\Gamma \left (\frac {1}{2}+\frac {r}{2}-k\right)} \frac {1}{g^{1+2k}} \, ,
\end{equation}
we obtain that the coefficients $b_n^{(3)}$ satisfy the following
(infinite) linear system
\begin{eqnarray}
b_{2n}^{(3)}&=& \sum _{m=0}^{n-1}(-1)^m 2^{m+\frac {1}{2}}\frac {B_{2m}}{(2m)!}b_{2n-2m+1}^{(3)} \, , \quad n \geq 1 \, , \nonumber  \\
&&  \label {b3eqs} \\
b_{2n+1}^{(3)}&=& \frac {\pi ^2 }{6 [\sigma ^{(0)}(0) ]^2} 4 (-1)^n
\frac {2^{n-\frac {1}{2}}-2^{\frac {1}{2}-n}}{(2n)!}B_{2n}+\sum
_{m=0}^{n}(-1)^m 2^{m+\frac {1}{2}}\frac
{B_{2m}}{(2m)!}b_{2n-2m+2}^{(3)} \, , \quad \quad n \geq 1 \, .
\nonumber
\end{eqnarray}
Again, the solution to these equations comes from the comparison
with the corresponding ones regarding the first generalised scaling
function \cite{FGR}, in the form of the simple mapping
\begin{equation}
b_{n}^{(3)}=\frac {\pi ^2 }{6 [\sigma ^{(0)}(0) ]^2} 2 b_{n-2}^{(1)}
\, , \quad n \geq 2 \, .
\end{equation}
Still the generating function
\begin{equation}
b^{(3)}(t)=\sum _{n=2}^{\infty} b_n^{(3)} t^n \, ,
\end{equation}
is simple, though, namely
\begin{equation}
b^{(3)}(t)=\frac {\pi ^2 }{6 [\sigma ^{(0)}(0) ]^2}\frac {2t^2}{\cos
\frac {t}{\sqrt {2}}-\sin \frac {t}{\sqrt {2}}} \, .
\end{equation}
This concludes what concerns the asymptotic part of the study about
$f_3(g)$. Instead, the non-perturbative (or non-analytic) terms will
be studied in subsection 3.4 (in comparison with those for
$f_4(g)$), as they give naturally rise to the limiting mass-gap.

\subsection{On the fourth and the other generalised scaling functions}

Thanks to the lesson of the previous sub-section, we are now in the
position to formulate a general scheme for computing the $n$-th
generalised scaling function $f_n(g)$ for $n \geq 2$ at arbitrary
value of the coupling constant. In particular, we can show how the
first computation, which is not reducible to the leading BES
equation, is exactly that for $f_4(g)$. Nevertheless, we will show
how the computation for $f_4(g)$ may be reduced to that for
$f_3(g)$, via a proportionality factor. The latter turns out to be
the only part which is not reducible to the BES equation. Actually,
similar situations would appear for the other scaling functions, as
may be guessed from the results of this sub-section.

In the limit (\ref{jlimit}) the scaling may be thought of as
governed by the function $S(k)$, defined through
\begin{equation}
S(k) \ln s= \frac {2\sinh \frac {|k|}{2}}{2\pi  |k|}\left [ \hat
\sigma _H (k)+ \hat \sigma _0^s (k) \right ] +\frac {e^{-\frac
{|k|}{2}}}{\pi |k|} \int  _{-\infty }^{\infty} \frac {dp}{2\pi}
\left [\hat \sigma _0^s(p)+\hat \sigma _H(p) \right ] \frac {\sin
(k-p)c} {k-p} \, , \label{S}
\end{equation}
and with the crucial expansion
\begin{equation}
S(k)= \sum _{n=0}^{\infty} S^{(n)}(k)j^n  \, \label{Sexpan}.
\end{equation}
Let us now concentrate our attention on $S^{(n)}(k)$, with $n\geq
2$, since the case $n=1$ is slightly different from the generic one
(as shown in \cite{FGR}). Restricting the domain to $k \geq 0$, we
expand again in Neumann's series (of Bessel functions)
\begin{equation}
S^{(n)}(k)=\sum _{p=1}^{\infty}S^{(n)}_{2p}(g)\frac {J_{2p}({\sqrt
{2}}gk)}{k}+\sum _{p=1}^{\infty}S^{(n)}_{2p-1}(g)\frac
{J_{2p-1}({\sqrt {2}}gk)}{k} \, ,
\end{equation}
and find the following equations for the coefficients of
$S^{(n)}(k)$, with $n\geq 2$,
\begin{eqnarray}
S^{(n)}_{2p}(g)&=&A_{2p}^{(n)}(g)-4p\sum _{m=1}^{\infty}Z_{2p,2m}(g)S^{(n)}_{2m}(g)+ 4p\sum _{m=1}^{\infty}Z_{2p,2m-1}(g)S^{(n)}_{2m-1}(g)  \nonumber \\
S^{(n)}_{2p-1}(g)&=&A_{2p-1}^{(n)}(g)-2(2p-1)\sum
_{m=1}^{\infty}Z_{2p-1,2m}(g)
S^{(n)}_{2m}(g)-\label{Seqn} \\
&-& 2(2p-1)\sum _{m=1}^{\infty}Z_{2p-1,2m-1}(g)S^{(n)}_{2m-1}(g)
\nonumber \, .
\end{eqnarray}
The forcing terms $A_r^{(n)}$ are given by the following integrals
\begin{equation}
A_r^{(n)}=r \int _{0}^{+\infty}\frac {dh}{2\pi h} \,  \frac
{J_{r}({\sqrt {2}}gh)}{\sinh \frac {h}{2}} \, \left. \int
_{-\infty}^{\infty} \frac {dp}{2\pi} 2 \frac {\sin (h-p)c }{h-p}
[\hat \sigma _0^s(p)+ \hat \sigma _H (p) ] \right |_{j^n} \, ,
\label {forcrn}
\end{equation}
where the symbol $|_{j^n}$ means that we wish to keep only the
coefficient of $j^n$ and also neglect its overall factor $\ln s$.

Thanks to (\ref{f_n}) and also by making use of (\ref{normale}), we
can generalise (\ref{Sf_3}) to
\begin{equation}
\lim _{k\rightarrow 0} S^{(n)}(k)=\frac {1}{2} f_n(g) .
\end{equation}
As before (cf. (\ref{f_3,2})), only the first component enters the
expression for the generalised constituent
\begin{equation}
f_n(g)={\sqrt {2}}g S_1^{(n)}(g).
\end{equation}

From the relations (\ref{normale}, \ref{c1,c2}), we
can gain the relevant power series expansion (here limited at the
order $j^4$ for simplicity's sake) \footnote{The term proportional
to $j$ in (\ref {forzexp}) needs careful consideration since it
receives additional contributions to produce the final inhomogeneous
equation of \cite{FGR}.}
\begin{eqnarray}
&& \int _{-\infty}^{\infty} \frac {dp}{2\pi} 2 \frac {\sin (h-p)c
}{h-p} [\hat \sigma _0^s(p)+ \hat \sigma _H (p)] =
\label {forzexp} \\
&=& \left [ -2\pi j + \frac {1}{3} \frac {\pi ^3}{[-4+\sigma
^{(0)}_H(0)]^2}h^2 j^3 - \frac {2}{3} \frac {\pi ^3[-4\ln 2 +\sigma
^{(1)}_H(0)]} {[-4+\sigma ^{(0)}_H(0)]^3}h^2 j^4+ O(j^5)\right ]\ln
s  \, . \nonumber
\end{eqnarray}
From direct inspection of the second and third term in the r.h.s. of
(\ref{forzexp}), we easily realise the proportionality between
$f_4(g)$ and $f_3(g)$:
\begin{equation}
f_4(g)=  -2 \frac {[-4\ln 2 +\sigma ^{(1)}_H(0)]} {[-4+\sigma
^{(0)}_H(0)]} f_3(g) \label{prop} \, ,
\end{equation}
which allows us to use the results about $f_3$, provided the non-BES
pre-factor be computed. In fact, this pre-factor contains explicitly
the first correction to the BES approach, i.e. $\sigma^{(1)}_H(0)$,
whose computation cannot apparently be derived from the BES
equation.

In general, it is thus possible to systematically expand the
solution (\ref{S}) into the form (\ref{Sexpan}-\ref{Seqn}), namely
order by order in $j$ up to the desired order. Albeit this procedure
seems to be in principle straightforward, it contains many intrigued
and intriguing details which are to be left for a dedicated
publication. Therefore, we plan to analyse in the next section the
non-analytic contributions in a numerical fashion (as we reckon it
quite effective for the current presentation), leaving the analytic
study about them for that publication.

\section{The non-analytic contributions from SYM and the $O(6)$ NLSM from string dual}
\setcounter{equation}{0}

Let us show how it is possible to make contact with the results of
the $O(6)$ NLSM \cite{DF}, which appears in the dual string theory
description, by analyzing the non-analytic exponential contributions
to $f_n(g)$.

First of all, let us focus our attention on $f_3(g)$. From the form
of the inhomogeneous term (\ref{A3}) (given by the $j^3$ coefficient
of (\ref{forzexp})), we may think of introducing a reduced form of
the equations (\ref{Seq2})
\begin{eqnarray}
S^{(3),red}_{2p}(g)&=&A^{(3),red}_{2p}(g)-4p\sum
_{m=1}^{\infty}Z_{2p,2m}(g)S^{(3),red}_{2m}(g)+
4p\sum _{m=1}^{\infty}Z_{2p,2m-1}(g)S^{(3),red}_{2m-1}(g)  \nonumber \\
S^{(3),red}_{2p-1}(g)&=&A^{(3),red}_{2p-1}(g)-2(2p-1)\sum
_{m=1}^{\infty}Z_{2p-1,2m}(g)
S^{(3),red}_{2m}(g)-  \label{3,red}\\
&-& 2(2p-1)\sum _{m=1}^{\infty}Z_{2p-1,2m-1}(g)S^{(3),red}_{2m-1}(g)
\nonumber \, ,
\end{eqnarray}
by defining a new inhomogeneous term
\begin{equation}
A_r^{(3),red}(g)=\frac{6 \, [-4+\sigma^{(0)}_H(0)]^2}{\pi
^2}A^{(3)}_r(g)= r \int _{0}^{+\infty}dh \, h \,  \frac
{J_{r}({\sqrt {2}}gh)}{\sinh \frac {h}{2}} \, ,
\end{equation}
which, thanks to linearity, easily entails
\begin{equation}
f_3(g) =  \frac {\pi ^2}{6 \, [-4+\sigma^{(0)}_H(0)]^2} f_3^{red}(g)
\, .
\end{equation}
Actually, the reduced system may be introduced for any $n$ of the
systems (\ref{Seqn}) and their peculiarity will be explained in a
future publication, as they admit a little involved mapping into the
BES equation (only as far as the computation of the first component
or $f_n(g)$ is concerned). On the contrary, the mapping could not be
found for the complete systems (\ref{Seqn}) (but, of course, for
$n=1,2,3$). Thanks to the proportionality (\ref{prop}), the reduced
part of $f_4(g)$ is still a very simple matter, as it virtually
coincides with $f_3(g)$, but the prefactor with $\sigma_H^{(1)}(0)$
could not be reformulated in terms of the BES equation: this makes
$f_4(g)$ the first (simple) general case.

Besides its analytical extra-value, the matrix inhomogeneous
(linear) system (\ref{3,red}) -- indeed equivalent to the linear
integral equation at this order -- may be explored numerically at
first quick instance, by truncating the dimension of the matrix
kernel (and vectors). Naturally, this treatment turns out into the
lines by \cite{BBKS} for the BES case, as it shares the same
(matrix) kernel: this strategy has been already efficiently
exploited in \cite{FGR} for the computation of the first mass-gap,
namely that related to $f_1(g)$. Thus, we obtain a new mass-gap
behaviour with the leading form
\begin{equation}
f_3^{red}(g) = k_2^{fit} g^{1/4} e^{-\frac{\pi}{\sqrt 2}g} \, .
\end{equation}
A fortiori, the numerical method \cite{BBKS} for the BES case in the
Fourier space may be directly applied to obtain the real space BES
estimate
\begin{equation}
\sigma^{(0)}_H(0) = 4 + k_1^{fit} g^{1/4} e^{-\frac{\pi}{\sqrt 2}g}
\, ,
\end{equation}
along with the best fit estimates for $k_1$ and $k_2$
\begin{equation}
k_1^{fit} = -7.1166 \pm 0.0005, \ \ \ \ \ k_2^{fit} = 5.5896 \pm
0.0005 \, .
\end{equation}
For clarity's sake, we should recall the exact mass-gap formula for
the $O(6)$ NLSM via a convergent Taylor series of the 't Hooft
coupling around $1/g=0$
\begin{equation}
m(g) =  k \, g^{1/4} e^{-\frac{\pi}{\sqrt 2}g} \left(
1+\frac{a_1}{g}+\frac{a_2}{g^2}+\dots \right) \,\,\, ,
\end{equation}
with the embedding pre-factor, $k$, fixed by the weak-coupling
perturbation theory (upon comparing versus the one-loop string
result \cite{FTT}) by \cite{AM} \footnote{All the other coefficients
of the series, i.e.  $a_1,\, a_2, \dots$\, become fixed by the
string UV embedding, in fact realised by the massive excitations.
The latter are also responsible for the quenched exponential terms,
$O(e^{-3\frac{\pi}{\sqrt 2}g})$, allowed by the SYM theory, and
instead forbidden by the NLSM.}
\begin{equation}
k= \frac{2^{5/8} \pi^{1/4}}{ \Gamma(5/4)} \,\,\,  .
\end{equation}
Eventually, we can easily realise how strongly the following exact
values for $k_1$, $k_2$ become suggested:
\begin{equation}
k_1 = - \pi k, \ \ \ \ \ k_2 = \frac{\pi^2}{4} k.
\end{equation}

To conclude the part on $f_3(g)$, we may write down its numerical
expression at strong coupling
\begin{equation}
f_3(g) = \frac{\pi^2}{6} \, (0.110366  \pm  0.000089) \, g^{-1/4}
e^{\frac{\pi}{\sqrt 2}g} \, ,
\end{equation}
or, alternatively, making use of the guessed exact values for $k_1$
and $k_2$, the exact (strong coupling) value
\begin{equation}
f_3(g) = \frac{\pi^2}{24 \, m(g)}.
\end{equation}
This is the same result as that derived by the exact expansion of
the $O(6)$ NLSM energy in the strong regime $j\ll m(g)$ as presented
in \cite{DF}, though this physics is explained uniquely by the free
theory.

Therefore, we ought to analyse in more detail the situation
regarding $f_4(g)$ to gain some physical insight into this peculiar
limit. For this purpose we shall evaluate the ratio $f_4(g)/f_3(g)$,
as coming from (\ref{prop}), at large values of $g$.

Let us first compute analytically the strong coupling limit of
$\sigma ^{(1)}_H(0)$ (indeed, a non-vanishing constant, as we will
see below). Simply using the definition of $S^{(1)}(k)$ as Neumann's
expansion with coefficients $S_m^{(1)}(g)$ \cite{FGR}, we obtain
\begin{equation}
\sigma ^{(1)}_H(0)=\int _{0}^{\infty} dk \frac {k}{\sinh \frac
{k}{2}} S^{(1)}(k)=\int _{0}^{\infty} dk \sum _{m=1}^{\infty}
S_m^{(1)}(g)\frac {J_m({\sqrt {2}}gk)}{\sinh \frac {k}{2}} \, .
\end{equation}
As we have recently found the exact asymptotic solution (around
$g=+\infty$) for the coefficients $S_m^{(1)}(g)$  \cite{FGR}, we
only need to sum over the index $m$ \footnote{Not to be confused
with the mass-gap $m(g)$.} so that
\begin{equation}
\sigma ^{(1)}_H(0)=\int _{0}^{\infty} \frac {dk}{\sinh \frac {k}{2}}
\left [ \sum _{n=1}^{\infty}b_{2n}^{(1)}(-1)^{n+1}\frac
{k^{2n}}{2^n} +
 \sum _{n=1}^{\infty}b_{2n-1}^{(1)}(-1)^{n}\frac {k^{2n-1}}
{2^{n-\frac {1}{2}}}\right ] \, .
\end{equation}
Eventually, we know from (4.4) of \cite{FGR} the generating function
$b^{(1)}(k)$ and thus we may integrate to obtain at leading order the
constant value
\begin{equation}
\sigma ^{(1)}_H(0)=2 \int _{0}^{\infty} dx \frac {1}{\sinh x} \left
[ 1-\frac {\cosh x}{\cosh 2x}-\frac {\sinh x}{\cosh 2x} \right ]=
3\ln 2 -\frac {\pi}{2} \, .
\end{equation}
On the other hand, we have already obtained above (cf. the
discussion concerning $f_3(g)$) the leading behaviour
\begin{equation}
-4 + \sigma ^{(0)}_H(0)=-\pi m(g) \, ,
\end{equation}
where $m(g)$ is exactly the $O(6)$ NLSM mass gap (the dependence on
the coupling $g$ by the density values in zero is omitted).
Therefore, we can conclude that
\begin{equation}
\frac {f_4(g)}{f_3(g)}= - \frac {2\ln 2 +\pi}{\pi m(g)} \, , \label
{f4/f3}
\end{equation}
which implies an exact relation for the first strong coupling term
\begin{equation}
f_4(g)=-\frac {\pi(2\ln 2 +\pi)}{24m^2(g)} \, . \label {f4result}
\end{equation}

This prediction is in remarkable agreement with the computation of
the ground state energy of the $O(6)$ NLSM  \cite{DF} and extends
the agreement between the latter model and the gauge theory up to
the first order ($j^4$) where the interaction goes on the stage. On
the contrary, the $O(6)$ NLSM approximation would no longer be
applicable for the next (exponential) order,
$O(e^{-3\frac{\pi}{\sqrt 2}g})$, present in the SYM theory with all
the subsequent ones, as it does not show up any terms of this form.
In fact, all the exponential corrections should be generated by the
massive (fermionic and bosonic) excitations of the dual string.
Hence, due to their origin, these corrections should be different in
the different $f_n(g)$: for the time being we can state the
difference at the next-to-leading order among the mass terms coming
from $f_1$, $f_3$ and $f_4$ in the gauge theory. A string theory
verification of this departure from the NLSM regime would be
desirable.

\section{Summary}

In this work we have initiated a general method for computing all
the generalised scaling function $f_n(g)$ appearing in the $sl(2)$
sector of ${\cal N}=4$ SYM when both the spin $s$ and the twist $L$
are very large, while the ratio $j=L/\ln s$ stays fixed (namely the
limit (\ref{jlimit})) . Our method relies on solving recursively the
linear equation (\ref{sigmaeq}) for the higher loop density of Bethe
roots $\sigma_H(u)$, which in its turn is completely determined by
the analogous solution of the one loop linear equation for the
density $\sigma_0(u)$. In particular, we have focused our attention
on the third and the fourth constituents to disentangle the
emergence, in the SYM theory, of different 'mass' terms, which all
flow to the mass-gap of the $O(6)$ Non-Linear Sigma Model:
comparison with the full string theory would be highly interesting,
though still missing. Very peculiarly, the convergence of the ground
state energy as a Taylor series in $j$ in the NLSM \cite{DF}
strongly suggests how its extension to the full theory, $f(g,j)=\sum
_{n=0}^{\infty} f_n(g)j^n$, should be convergent too. In conclusion,
we have left apart some details about the analytic calculations and
the general structure, bearing in mind a more complete and
systematic study for the near future.

\medskip

{\bf Acknowledgments} We thank F. Buccheri, D. Bombardelli and F.
Ravanini for many intriguing suggestions, the INFN grant "Iniziativa
specifica PI14" and the international agreement INFN-MEC-2008 for
travel financial support. P.G. work is partially supported by
MEC-FEDER (grant FPA2005-00188), by the Spanish Consolider-Ingenio
2010 Programme CPAN (CSD2007-00042) and by Xunta de Galicia
(Conseller\'\i a de Educaci\'on and grant PGIDIT06PXIB296182PR).

\end{document}